\documentclass{JHEP3} 
\usepackage{amssymb,amsmath,amsthm,bbm}

\newcommand{\eq}[1] 
  {\begin{equation} 
    #1 
  \end{equation}}

\newcommand{\spliteq}[1] 
  {\begin{eqnarray} 
    #1 
  \end{eqnarray}}

\newcommand{\aligneq}[1] 
  {\begin{align} 
    #1 
  \end{align}}

\newcommand{\Res}{{\rm Res}}
\newcommand{\brac}[1]{\left(#1\right)} 
\newcommand{\set}[1]{\left\{#1\right\}}
\newcommand{\ebrac}[1]{\left[#1\right]} 
\newcommand{\abs}[1]{\left\arrowvert #1\right\arrowvert}

\newcommand{\cor}[1]{\left<#1\right>}

 \newcommand{\sgn}{\mathrm{sgn}}

\title{Continuously Crossing $\bf u=z$ in the $\bf H_3^+$ Boundary CFT}

\author{Hendrik Adorf and Michael Flohr\\ Institut f\"ur Theoretische Physik,\\ Gottfried Wilhelm
Leibniz Universit\"at Hannover,\\ Appelstra\ss e 2, 30167 Hannover, Germany.\\ E-mail:
\email{adorf, flohr@itp.uni-hannover.de}}

\preprint{ITP--UH-14/07}

\abstract{For $AdS$ boundary conditions, we give a solution of the $\rm H_3^+$ two point function
involving degenerate field with ${\rm SL}(2)$-label $b^{-2}/2$, which is defined on the full
$(u,z)$ unit square. It consists of two patches, one for $z<u$ and one for $u<z$. Along the $u=z$
"singularity", the solutions from both patches are shown to have finite limits and are merged
continuously as suggested by the work of Hosomichi and Ribault. From this two point function, we
can derive $b^{-2}/2$-shift equations for $AdS_2$ D-branes. We show that discrete as well as
continuous $AdS_2$ branes are consistent with our novel shift equations without any new
restrictions.}


\begin{document} \section{\label{Intro}Introduction} In the study of non-compact and non-rational
conformal field theories (CFTs), the $\rm H_3^+$ model (besides Liouville theory) serves as a
basic tractable example. Accordingly, hope is raised that it will teach us some important lessons
about the general features of this class of CFTs. One of these lessons, which has been discussed
in \cite{HosomichiRibault:SolutionOnDisc} and which becomes important in the boundary theory of
the $\rm H_3^+$ model, is the weakening of the Cardy-Lewellen constraints. This lesson shall be
taken up in the present paper.

A possible approach to the boundary $\rm H_3^+$ CFT is to construct two point functions involving
a degenerate field as solutions of Knizhnik-Zamolodchikov equations. This is useful, because by
taking a certain factorization limit, these special two point functions allow to derive shift
equations that restrict the one point functions of the theory. Usually these constraints can be
solved and the one point functions obtained. In case of degenerate field with ${\rm SL}(2)$-label
$j=1/2$ (please refer to section \ref{Conventions} for an introductory overview of the $\rm
H_3^+$ model) this procedure has succesfully been carried out in \cite{PST}. But since that
solution is not unique (for example, multiplication with an arbitrary $1/2$-periodic function
again yields a solution), a further shift equation would be desirable. For its derivation, it is
most convenient to use the next simple degenerate field, which has ${\rm SL}(2)$-label
$b^{-2}/2$.

For that degenerate field however there are some difficulties in constructing the two point
function in a region of the $(u,z)$-plane that also covers the domain in which the factorization
limit is to be taken. While a solution to the Knizhnik-Zamolodchikov equation can be given in the
region $z<u$, it was unclear up to now how it could be continued to the patch $u<z$, which is the
patch relevant to the factorization limit. In particluar, a suitable continuation prescription is
needed.

Such a prescription has been suggested in \cite{HosomichiRibault:SolutionOnDisc} by Hosomichi and
Ribault.\footnote{We like to mention, however, that a different prescription has also been
analysed in \cite{Adorf:VariousH3Branes}.} They study a mapping of $\rm H_3^+$ to Liouville
correlators. This mapping is formulated in two different regimes: The bulk regime and the
boundary regime. These two regimes do not have any overlap and therefore, the Cardy-Lewellen
constraints have to be supplemented by a further requirement. Hosomichi and Ribault demand that
all $\rm H_3^+$ correlators be continuous when changing from one regime into the other. In our
setting, bulk and boundary regime correspond to the patches $z<u$ and $u<z$ respectively.

With that motivation, we now set out to construct the $AdS_2$ boundary two point function involving
degenerate field with ${\rm SL}(2)$-label $b^{-2}/2$ in the following way: After introducing our
notation conventions, we first give a solution to the Knizhnik-Zamolodchikov equation in the
region $z<u$. It is fixed from the asymptotics of an operator product expansion (OPE). We show
that this solution has a finite $u=z$ limit. Then, a solution to the Knizhnik-Zamolodchikov
equation in the region $u<z$ is found. It is partially fixed from the requirement that its $u=z$
limit matches that of the previous solution. However, an ambiguity in the conformal blocks ${\cal
F}^{s}_{j,-}$ and ${\cal F}^{s}_{j,\times}$ persists. Yet, the two point function is then defined
everywhere in the $(u,z)$ unit square and continuous along $u=z$. This construction is the
content of section \ref{ConstructionTwoPoint}. Afterwards, in section \ref{Limit}, we take the
factorization limit and derive the desired $b^{-2}/2$-shift equations for discrete as well as
continuous $AdS_2$ D-branes. They constitute a new and independent constraint. The key point is
really that the aforementioned ambiguity does not enter here, because the conformal blocks ${\cal
F}^{s}_{j,-}$ and ${\cal F}^{s}_{j,\times}$ are shown not to contribute in the factorization
limit. In section \ref{Consistency}, we finally check that discrete as well as continuous $AdS_2$
branes are consistent with our new shift equations. We discuss our results in the light of
\cite{HosomichiRibault:SolutionOnDisc} and Cardy's work \cite{Cardy:BCsFusionAndVerlinde}.

\section{\label{Conventions}Bulk and Boundary $\bf H_3^+$ CFT - Notation and Conventions} Our
notation should coincide almost everywhere with that used in standard references like \cite{PST}
or \cite{Teschner:StructureConstantsAndFusionRules}. The collected facts of this chapter can also
all be found there.

Due to an affine $\hat{\rm sl}(2,\mathbb{C})_k\times\hat{\rm sl}(2,\mathbb{C})_k$ symmetry, the
primary fields in the $\rm H_3^+$ CFT are organized in ${\rm sl}(2,\mathbb{C})$ representations
and thus labelled by a pair of ${\rm sl}(2,\mathbb{C})$ 'spins' $(j,\bar{j})$, as well as a pair
of internal variables, which we will denote $(u,\bar{u})\in\mathbb{C}^2$. Concentrating on one
chiral half only, we write a primary field as $\Theta_{j}(u|z)$, with worldsheet coordinate
$z\in\mathbb{C}$. The $\hat{\rm sl}(2,\mathbb{C})_k$ currents $J^a(z)$ act via the following
operator product expansion (OPE) 
\eq{J^a(z)\Theta_j(u|w)=\frac{D^a_j(u)\Theta_j(u|w)}{z-w},} i.e.
the zero mode algebra is represented through differential operators $D^a_j(u)$, given by
\eq{D^+_j(u):=-u^2\partial_u+2ju,\qquad D^-_j(u):=\partial_u,\qquad D^3_j(u):=u\partial_u-j\,.}
As usual, the Sugawara construction expresses the energy momentum tensor of the theory in terms of products of the currents and thereby establishes the following relation between conformal weight $h$ and 'spin'-label $j$ of primary fields: 
\eq{h\equiv h(j)=-\frac{j(j+1)}{k-2}=:-b^2 j(j+1)\,.} 
It is important to note the reflection symmetry $h(-j-1)=h(j)$, which enables one to identify the
representations with labels $j$ and $-j-1$. The corresponding primary fields $\Theta_j(u|z)$ and
$\Theta_{-j-1}(u|z)$ are then related as:
\eq{\label{RefSymm}\Theta_j(u|z)=-R(-j-1)\frac{2j+1}{\pi}\int_{\mathbb{C}}{\rm d}^2
u'|u-u'|^{4j}\Theta_{-j-1}(u'|z)\,,} 
whith the following expression for the reflection amplitude $R(j)$: 
\eq{\label{Rj}R(j)=-\nu_b^{2j+1}\frac{\Gamma(1+b^2(2j+1))}{\Gamma(1-b^2(2j+1))}\,,} 
and $\nu_{b}=\pi\frac{\Gamma(1-b^2)}{\Gamma(1+b^2)}$. The physical spectrum (normalizable operators) consists of the continuous ${\rm sl}(2,\mathbb{C})$ representations \cite{Teschner:OPE}, that are parametrized by $j\in -\frac{1}{2}+i\mathbb{R}_{>0}$. They are infinite dimensional representations.\\

\noindent So far for the bulk theory. Now, introducing a boundary amounts to considering the
model on (the closure of) the upper half plane $z\in\bar{\mathbb{H}}:=\set{z\in\mathbb{C}|{\rm
Im}(z)\geq 0}$ with suitable boundary conditions along the real axis. The maximal symmetry
preserving boundary conditions are obtained by imposing a so-called gluing condition
\eq{\label{GlueCond}J^a(z)=\rho^a_{\hphantom{a}b}\bar{J}^b(\bar{z}) \qquad {\rm at}\hskip .1cm z=\bar{z}}
with gluing map $\rho$. $\rho$ is an automorphism of the current algebra which leaves the energy
momentum tensor invariant. Thus, we also have 
\eq{\label{TGlue}T(z)=\bar{T}(\bar{z}) \qquad {\rm at}\hskip .1cm z=\bar{z}\,.} 
The meaning of (\ref{GlueCond}) and (\ref{TGlue}) is, that besides a subgroup of the current algebra symmetry, also half of the conformal symmetry is preserved. For the purposes of the present paper, we will only deal with the gluing map
\eq{\rho\bar{J}^3=\bar{J}^3, \qquad \rho\bar{J}^{\pm}=-\bar{J}^{\pm}} 
and the associated branes are conventionally called $AdS_2$ D-branes. The conformal Ward identites fix the one point function in the presence of $AdS_2$ boundary condition $\alpha\in\mathbb{R}$ to be of the form
\eq{\cor{\Theta_j(u|z)}_{\alpha}=\abs{z-\bar{z}}^{-2h(j)}\abs{u+\bar{u}}^{2j}
A_{\sigma}(j|\alpha)\,.} 
We call the
unknown function $A_{\sigma}(j\arrowvert\alpha)$ the one point amplitude. Note that it still
depends on $\sigma:=\sgn(u+\bar{u})$. It is interpreted as the strength of coupling of a closed
string with label $j$ to the brane labelled by $\alpha$. The strategy of this paper will be to
derive necessary conditions on $A_{\sigma}(j\arrowvert \alpha)$ by considering two point
functions involving a degenerate field (section \ref{ConstructionTwoPoint}) and then taking a
factorization limit (section \ref{Limit}). 

One important constraint on the one point amplitude can already be stated here. It stems from the reflection symmetry (\ref{RefSymm}) and for our choice of boundary conditions (\ref{GlueCond}) reads:
\aligneq{\frac{\pi}{2j+1}&\abs{u+\bar{u}}^{2j}A_{\sigma}(j|\alpha)=\nonumber\\
&=-R(-j-1)\int_{\mathbb{C}}d^2u'\abs{u-u'}^{4j}\abs{u'+\bar{u}'}^{-2j-2}
A_{\sigma'}(-j-1|\alpha)\,.} 
Expanding $A_{\sigma'}(-j-1\arrowvert\alpha)=A^0(-j-1\arrowvert\alpha)+\sigma'
A^1(-j-1\arrowvert\alpha)$, we are lead to compute the occuring integral ($\epsilon\in\set{0,1}$):
\eq{I_{\epsilon}:=\int_{\mathbb{C}}d^2u'\abs{u-u'}^{4j}\abs{u'+\bar{u}'}^{-2j-2}
(\sigma')^{\epsilon}\,.} 
It can be carried out by elementary means. We obtain
\eq{I_{\epsilon}=-\frac{\pi}{2j+1}\abs{u+\bar{u}}^{2j}(-\sigma)^{\epsilon}\,.}
Hence, the reflection symmetry constraint becomes
\eq{A_{\sigma}(j\arrowvert\alpha)=R(-j-1)A_{-\sigma}(-j-1\arrowvert\alpha)\,.} 
For later purposes, we like to introduce a redefinition of the one point amplitude here. It is motivated by the form of the reflection symmetry constraint just written down. Namley, defining
\eq{\label{ReDef}f_{\sigma}(j):=\nu _b^j\Gamma(1+b^2(2j+1))A_{\sigma}(j\arrowvert\alpha)} 
(note that we have dropped the $\alpha$-dependence of $f_{\sigma}$), it is easy to see, using equation (\ref{Rj}) for $R(j)$, that we now simply have
\eq{\label{RefSymmCnstr}f_{\sigma}(j)=-f_{-\sigma}(-j-1)\,.}

\section{\label{ConstructionTwoPoint}Construction of the Two Point Function} From the Ward
identities of the model, the two point function 
\eq{G^{(2)}_{j,\alpha}(u_i\arrowvert z_i):=\cor{\Theta_{b^{-2}/2}(u_2\arrowvert z_2)\Theta_j(u_1\arrowvert z_1)}_{\alpha}} 
is restricted to be of the form 
\spliteq{\label{B2pt}G^{(2)}_{j,\alpha}(u_1,u_2|z_1,z_2)&=&\abs{z_1-\bar{z}_1}^{2[h(b^{-2}/2)-h(j)]}\abs{z_1-\bar{z}_2}^{-4h(b^{-2}/2)}\times\nonumber\\
&&\times\abs{u_1+\bar{u}_1}^{2j-b^{-2}}\abs{u_1+\bar{u}_2}^{2b^{-2}}H^{(2)}_{j,\alpha}(u\arrowvert z)\,.} 
The parameter $\alpha$ again labels the $AdS_2$ boundary conditions. The reduced two point function $H^{(2)}_{j,\alpha}(u\arrowvert z)$ is a still unknown function of the crossing ratios \eq{z:=\frac{\abs{z_2-z_1}^2}{\abs{z_2-\bar{z}_1}^2} \hskip .5cm {\rm and} \hskip
.5cm u:=\frac{\abs{u_2-u_1}^2}{\abs{u_2+\bar{u}_1}^2}\,.} 
The two point function (\ref{B2pt}) has to satisfy a Knizhnik-Zamolodchikov equation. For the coordinate $z_2$ this equation reads
\aligneq{-\frac{1}{b^2}\partial_{z_2}G^{(2)}_{j,\alpha}(u_i\arrowvert z_i)=&\nonumber\\ 
=\sum_a {\cal D}_{b^{-2}/2}^a(u_2)&\otimes\ebrac{\frac{{\cal D}_j^a(u_1)}{z_2-z_1}+\frac{\rho^a_{\hphantom{a}b}\bar{{\cal
D}}_j^b(\bar{u}_1)}{z_2-\bar{z}_1}+\frac{\rho^a_{\hphantom{a}b}\bar{{\cal
D}}_{b^{-2}/2}^b(\bar{u}_2)}{z_2-\bar{z}_2}}G^{(2)}_{j,\alpha}(u_i\arrowvert z_i)\,.}
Mapping $z_1\rightarrow 0$, $\bar{z}_2\rightarrow 1$ and $\bar{z}_1\rightarrow\infty$ (i.e.
$z_2\rightarrow z$), it is brought to standard form
\aligneq{\label{KniZaStd}-&b^{-2}z(z-1)\partial_{z}H^{(2)}_{j,\alpha}(u\arrowvert
z)=u(u-1)(u-z)\partial^2_u H^{(2)}_{j,\alpha}+\nonumber\\
&+\set{\ebrac{1-2b^{-2}}u^2+\ebrac{b^{-2}-2j-2}uz+\ebrac{2j+b^{-2}}u+z}\partial_u
H^{(2)}_{j,\alpha}+\nonumber\\
&+\set{b^{-4}u+\ebrac{b^{-2}j-b^{-4}/2}z-b^{-2}j}H^{(2)}_{j,\alpha}\,.} 
This is solved by (see \cite{Appell} and also \cite{Teschner:StructureConstantsAndFusionRules}\footnote{The solution given in \cite{Teschner:StructureConstantsAndFusionRules} is slightly different and exists in a
smaller domain of the $(u,z)$-plane. It does however coincide with the solution given here on the
overlap of domains of existence}) $H^{(2)}_{j,\alpha}=\sum_{\epsilon=+,-,\times}
a^{j}_{\epsilon}(\alpha){\cal F}^{s}_{j,\epsilon}$ with 
\spliteq{\label{ConfBlocks1+}{\cal
F}^{s}_{j,+}(u\arrowvert z)&=&z^{-j}(1-z)^{-b^{-2}/2}F_1(\alpha,\beta,\beta';\gamma\arrowvert
u;z)\,,\\ 
\label{ConfBlocks1-}{\cal F}^{s}_{j,-}(u\arrowvert z)&=&z^{-j}(1-z)^{-b^{-2}/2}u^{-\beta}z^{1+\beta-\gamma}\times\nonumber\\ 
&&\times F_1\left(1+\beta+\beta'-\gamma,\beta,1+\alpha-\gamma;2+\beta-\gamma\left\arrowvert\frac{z}{u};z
\right)\right.,\\ 
\label{ConfBlocks1x}{\cal F}^{s}_{j,\times}(u\arrowvert z)&=&z^{-j}(1-z)^{-b^{-2}/2}u^{1-\gamma}\times\nonumber\\ 
&&\times G_2\left(\beta',1+\alpha-\gamma;1+\beta-\gamma,\gamma-1\left\arrowvert
-\frac{z}{u};-u\right.\right).} 
The appearance of only three conformal blocks is due to the presence of degenerate field $\Theta_{b^{-2}/2}$. The propagating modes are denoted $j_{\pm}:=j\pm b^{-2}/2$ and $j_{\times}:=-j-1-b^{-2}/2$. We identify the parameters to be
\eq{\label{Parameters}\alpha=\beta=-b^{-2},\hskip .3cm \beta'=-2j-1-b^{-2},\hskip .3cm
\gamma=-2j-b^{-2}\,.} 
Splitting the common factor $z^{-j}(1-z)^{-b^{-2}/2}$, these functions are found in \cite{Appell} as (respectively) ${\cal Z}_1$, ${\cal Z}_{15}$ and the last one is related to ${\cal Z}_8$. The functions $F_1$ and $G_2$ are generalized hypergeometric functions: $F_1$ is the first one of Appell's double hypergeometric functions (see \cite{Appell}, \cite{Bateman}, \cite{Exton} for more information). The function $G_2$ is one of Horn's functions (see for example \cite{Bateman} and \cite{Exton}). We give their definitions as convergent series and some of their properties in the appendix. The relation between ${\cal Z}_8$ and (\ref{ConfBlocks1x}) is as follows: By analytically continuing ${\cal Z}_8$ to the domain around $(\infty,0)$, a sum of the function ${\cal Z}_1$ and the above $u^{1-\gamma}G_2$ is produced. Therefore, since ${\cal Z}_1$ solves Appell's differential equation, so does $u^{1-\gamma}G_2$. (\ref{ConfBlocks1+})-(\ref{ConfBlocks1x}) constitute a linearly independent set of three solutions. By general theory, any other solution can be expressed as a linear combination of them \cite{Exton}. This reflects nicely the fact that the degenerate field $\Theta_{b^{-2}/2}$ restricts the propagating fields to only three possibilities, namely those belonging to representations $j_{\pm}$ and $j_{\times}$, as we have mentioned above. 

The conformal blocks (\ref{ConfBlocks1+}), (\ref{ConfBlocks1-}), (\ref{ConfBlocks1x}) are
obviously well defined in the patch $z<u$ (when talking about the patches, it is always tacitly
understood that $0\leq u<1$ and $0\leq z<1$). Their linear combinations, i.e. the coefficients
$a^{j}_{\epsilon}(\alpha)$, are determined from comparison with the OPE in the limit
$z\rightarrow 0$ followed by $u\rightarrow 0$. This has been discussed in
\cite{Teschner:StructureConstantsAndFusionRules}. The result is simply
\eq{\label{LinComb}a^{j}_{\epsilon}(\alpha)=C_{\epsilon}(j)
A_{\sigma}(j_{\epsilon}|\alpha)\,,} 
$C_{\epsilon}(j)$ being the coefficients occuring in the OPE of $\Theta_{b^{-2}/2}(u_2|z_2)$ with $\Theta_j(u_1|z_1)$. They are given in appendix \ref{OPE}.\\

\noindent Let us now see how this solution can be extended to the region $u<z$. Clearly, ${\cal
F}^{s}_{j,+}$ is already everywhere defined, so we do not have to worry about it in the
following. But let us analyse how ${\cal F}^{s}_{j,-}$ and ${\cal F}^{s}_{j,\times}$ behave when
we move to $u=z$ from the region $z<u$. Using the generalized series representations of $F_1$ and
$G_2$ (see appendix \ref{AppellAndHorn}), we find 
\spliteq{{\cal F}^{s}_{j,-}(u=z)&=&z^{1-\gamma-j}(1-z)^{-b^{-2}/2}\frac{\Gamma(1-\beta-\beta')\Gamma(2+\beta-\gamma)}{\Gamma(1-\beta')\Gamma(2-\gamma)}\times\nonumber\\ &&\times
F(1+\beta+\beta'-\gamma,1+\alpha-\gamma;2-\gamma\arrowvert z)\,,\\ {\cal
F}^{s}_{j,\times}(u=z)&=&z^{1-\gamma-j}(1-z)^{-b^{-2}/2}\frac{\Gamma(1-\beta-\beta')\Gamma(\gamma
-\beta)}{\Gamma(1-\beta)\Gamma(\gamma-\beta-\beta')}\times\nonumber\\ &&\times
F(1+\beta+\beta'-\gamma,1+\alpha-\gamma;2-\gamma\arrowvert z)\,.} 
Here, $F$ denotes the standard hypergeometric function. Interestingly, the linearly independent solutions (\ref{ConfBlocks1-}), (\ref{ConfBlocks1x}) degenerate at $u=z$ and become essentially the same function (up to factors). We will see shortly that it is this fact that prevents us from fixing a solution for $u<z$ uniquely.

The task is now to find a solution to the Knizhnik-Zamolodchikov equation in the region $u<z$
that matches the above for $u=z$. One building block is, of course, ${\cal F}^{s}_{j,+}$. The two
others are 
\spliteq{\label{ConfBlocks2-}\tilde{{\cal F}}^{s}_{j,-}(u\arrowvert
z)&=&z^{-j}(1-z)^{-b^{-2}/2}u^{1+\beta'-\gamma}z^{-\beta'}\times\nonumber\\ &&\times
F_1\left(1+\beta+\beta'-\gamma,1+\alpha-\gamma,\beta';2+\beta'-\gamma\left\arrowvert
u;\frac{u}{z}\right.\right),\\ \tilde{\label{ConfBlocks2x}{\cal F}}^{s}_{j,\times}(u\arrowvert
z)&=&z^{-j}(1-z)^{-b^{-2}/2}z^{1-\gamma}\times\nonumber\\ &&\times
G_2\left(\beta,1+\alpha-\gamma;1+\beta'-\gamma,\gamma-1\left\arrowvert
-\frac{u}{z};-z\right.\right).} 
The tilde indicates that this is the solution in region $u<z$. Again, splitting the common factor $z^{-j}(1-z)^{b^{-2}/2}$, the first function is found in \cite{Appell} as ${\cal Z}_{14}$ and the second one is related to ${\cal Z}_9$ in a similar manner as before. Note that the third argument of $G_2$ is $1+\beta'-\gamma=0$ for our specific parameter values (\ref{Parameters}) which are dictated by the Knizhnik-Zamolodchikov equation. Nevertheless, the function $G_2$ stays well-defined and a generalized series representation can be derived (see appendix \ref{AppellAndHorn}). By making use of the general series representations of $F_1$ and $G_2$, one can show that the conformal blocks (\ref{ConfBlocks2-}), (\ref{ConfBlocks2x}) agree along $u=z$ with those from patch $z<u$ up to factors:
\spliteq{\tilde{{\cal
F}}^{s}_{j,-}(u=z)&=&z^{1-\gamma-j}(1-z)^{-b^{-2}/2}\frac{\Gamma(1-\beta-\beta')\Gamma(2+\beta'-\gamma)}{\Gamma(1-\beta)\Gamma(2-\gamma)}\times\nonumber\\ &&\times
F(1+\beta+\beta'-\gamma,1+\alpha-\gamma;2-\gamma\arrowvert z)\,,\\ \tilde{{\cal
F}}^{s}_{j,\times}(u=z)&=&z^{1-\gamma-j}(1-z)^{-b^{-2}/2}\frac{\Gamma(2-\beta-\gamma)}{\Gamma(1-\beta)\Gamma(2-\gamma)}\times\nonumber\\ &&\times
F(1+\beta+\beta'-\gamma,1+\alpha-\gamma;2-\gamma\arrowvert z)\,.} 
These factors are absorbed through a suitable definition of the expansion coefficients $\tilde{a}^{j}_{\epsilon}(\alpha)$ in the patch $u<z$. They must therefore be related to the former ones $a^{j}_{\epsilon}(\alpha)$ as
\begin{gather} \label{LinComb2+}\tilde{a}^{j}_{+}(\alpha)=a^{j}_{+}(\alpha)\,,\\
\begin{split}\label{LinComb2-x}&\tilde{a}^{j}_{-}(\alpha)\frac{\Gamma(1-\beta-\beta')\Gamma(2+\beta'-\gamma)}{\Gamma(1-\beta)\Gamma(2-\gamma)}+\tilde{a}^{j}_{\times}(\alpha)\frac{\Gamma(2-\beta-\gamma)}{\Gamma(1-\beta)\Gamma(2-\gamma)}=\\
&\phantom{\tilde{a}^{j}_{-}(\alpha)}=a^{j}_{-}(\alpha)\frac{\Gamma(1-\beta-\beta')\Gamma(2+\beta-
\gamma)}{\Gamma(1-\beta')\Gamma(2-\gamma)}+a^{j}_{\times}(\alpha)\frac{\Gamma(1-\beta-\beta')\Gamma(\gamma-\beta)}{\Gamma(1-\beta)\Gamma(\gamma-\beta-\beta')}\,.
\end{split} 
\end{gather} 
Thus, we cannot uniquely fix the coefficients $\tilde{a}^{j}_{-}(\alpha)$ and
$\tilde{a}^{j}_{\times}(\alpha)$. An ambiguity remains in the two dimensional subspace spanned by
$\tilde{{\cal F}}^{s}_{j,-}$ and $\tilde{{\cal F}}^{s}_{j,\times}$. It is good to realize, that
for the values of the parameters $\alpha$, $\beta$, $\beta'$, $\gamma$ which are given in
(\ref{Parameters}) and ${\rm SL}(2)$-label $j$ in the physical range $j\in -\frac{1}{2}+{\rm
i}\mathbb{R}_{>0}$, we never catch any poles of the gamma functions. The reduced two point
function $H^{(2)}_{j,\alpha}=\sum_{\epsilon=+,-,\times} a^{j}_{\epsilon}(\alpha){\cal
F}^{s}_{j,\epsilon}$ is now defined in the (semi-open) unit square $0\leq u<1$, $0\leq z<1$. The
lines $u=1$, $z=1$ have to be understood as limiting cases.

\section{\label{Limit}Factorization Limit and Shift Equations} 
Using our solution (\ref{ConfBlocks1+}), (\ref{ConfBlocks2-}), (\ref{ConfBlocks2x}) in the patch $u<z$, we can now take the limit $z\rightarrow 1$ from below while $u<1$. Performing it on the conformal blocks, we find 
\spliteq{\label{PlusLim}\tilde{{\cal F}}^{s}_{j,+}&\simeq&(1-z)^{1+b^{-2}/2}(1-u)^{b^{-2}}\frac{\Gamma(\gamma)\Gamma(\alpha+\beta'-\gamma)}{\Gamma(\alpha)\Gamma(\beta')}\cdot\ebrac{1+{\cal O}(1-z)}+\nonumber\\
&&+(1-z)^{-b^{-2}/2}\frac{\Gamma(\gamma)\Gamma(\gamma-\alpha-\beta')}{\Gamma(\gamma-\alpha)\Gamma
(\gamma-\beta')}F(\alpha,\beta;\gamma-\beta'\arrowvert u)\cdot\ebrac{1+{\cal O}(1-z)}\,,\\
\tilde{{\cal F}}^{s}_{j,-}&\simeq&(1-z)^{-b^{-2}/2}u^{1+\beta'-\gamma}\times\nonumber\\ &&\times
F(1+\beta+\beta'-\gamma,1+\alpha+\beta'-\gamma;2+\beta'-\gamma\arrowvert u)\cdot\ebrac{1+{\cal
O}(1-z)}\,,\\ \tilde{{\cal
F}}^{s}_{j,\times}&\simeq&(1-z)^{-b^{-2}/2}F(\alpha,\beta;1|u)\ebrac{1+{\cal O}(1-z)}\,.} 
The limit $z\rightarrow 1$ from below corresponds to using a bulk-boundary OPE in the correlator.
Now, there are two cases to distinguish, as is explained in detail in \cite{Schomerus:Lectures2}: Assuming a discrete open string spectrum on the brane, the bulk-boundary OPE for $\Theta_{b^{-2}/2}$ is
\spliteq{\Theta_{b^{-2}/2}(u_2\arrowvert
z_2)&=&\abs{z_2-\bar{z}_2}^{1+b^{-2}/2}\abs{u_2+\bar{u}_2}^{b^{-2}}C_{\sigma}(b^{-2}/2,0|\alpha)\mathbbm{1}\set{1+{\cal O}\brac{z_2-\bar{z}_2}}+\nonumber\\
&+&\abs{z_2-\bar{z}_2}^{-b^{-2}/2}\abs{u_2+\bar{u}_2}^{2b^{-2}+1}C_{\sigma}(b^{-2}/2,b^{-2}|\alpha)\times\nonumber\\
&&\hphantom{+\abs{z_2-\bar{z}_2}^{-b^{-2}/2}}\times
\brac{{\cal J}\Psi}^{\alpha\,\alpha}_{b^{-2}}\brac{u_2\left|{\rm Re}(z)\right.}\set{1+{\cal O}\brac{z_2-\bar{z}_2}}+\nonumber\\
&+&\abs{z_2-\bar{z}_2}^{-b^{-2}/2}C_{\sigma}(b^{-2}/2,-b^{-2}-1|\alpha)\times\nonumber\\
&&\hphantom{+\abs{z_2-\bar{z}_2}^{-b^{-2}/2}}\times
\brac{{\cal J}\Psi}^{\alpha\,\alpha}_{-b^{-2}-1}\brac{u_2\left|{\rm Re}(z)\right.}\set{1+{\cal O}\brac{z_2-\bar{z}_2}}\,,}
where we have defined 
\eq{({\cal J}\Psi)^{\alpha\,\alpha}_{l}(u|z):=\int_{\mathbb{R}}\frac{dt}{2\pi} \abs{u+it}^{-2l-2}\Psi^{\alpha\,\alpha}_{l}(t|z)\,.}
For the purpose of deriving the factorization constraint, we concentrate on the contribution of the identity field $\mathbbm{1}$ only. Identifying $C_{\sigma}(b^{-2}/2,0|\alpha)=A_{\sigma}(b^{-2}/2|\alpha)$, we deduce the following $b^{-2}/2$-shift equation
\eq{\label{DiscreteShiftEq}f_{\sigma}(b^{-2}/2)f_{\sigma}(j)=\Gamma(1+b^2)
f_{\sigma}(j+b^{-2}/2)\,,} 
where we have suppressed the $\alpha$-dependence and used the redefined one point amplitude
(\ref{ReDef}). Note that on the LHS, the one point amplitudes carry identical $\sigma$'s. This is
because we are in a region where $u<1$. In a domain with $1<u$ they would indeed carry opposite
signs.

On the other hand, assuming a continuous open string spectrum on the brane, the bulk-bundary OPE of $\Theta_{b^{-2}/2}$ contains 
\eq{\tilde{c}_{\sigma}(b^{-2}/2,j_{\epsilon}|\alpha):=\Res_{j_2=b^{-2}/2}C_{\sigma}(j_2,j_{\epsilon}|\alpha)} 
instead of $C(b^{-2}/2,j_{\epsilon}|\alpha)$ (as usual, $\epsilon=+,-,\times$). The reason for this is given in \cite{Schomerus:Lectures2}. Let us summarize it here briefly: Since we are using Teschner's Trick, i.e. we are analytically continuing the field label $j_2$ to the label of a degenerate representation (which is here $j_2=b^{-2}/2$), we should look at the generic bulk-boundary OPE
\eq{\Theta_{j_2}(u_2|z_2)\simeq\int_{{\cal C}^+}dl\abs{z_2-\bar{z}_2}^{-2h(j_2)+h(l)}
\abs{u_2+\bar{u}_2}^{2j_2+l+1}C_{\sigma}(j_2,l|\alpha)\brac{{\cal J}\Psi}^{\alpha\,\alpha}_{l}\brac{u_2\left|{\rm Re}(z_2)\right.},}
where the contour of integration is ${\cal C}^+:=-\frac{1}{2}+i\mathbb{R}$. Since $j_2=b^{-2}/2$ is a degenerate representation, only a discrete set of open string modes is excited in the bulk-boundary OPE of its corresponding field operator. Accordingly, when deforming the contour in the process of analytic continuation, only finitely many contributions are picked up. They come from poles that develop in the $C_{\sigma}(j_2,l|\alpha)$. Therefore, not the bulk-boundary coefficients themselves, but only their residua occur. Focussing on the identity channel again, we obtain
\eq{\Theta_{b^{-2}/2}(u_2|z_2)\simeq\abs{z_2-\bar{z}_2}^{1+b^{-2}/2}\abs{u_2+\bar{u}_2}^{b^{-2}+1}\tilde{c}_{\sigma}(b^{-2}/2,0|\alpha)\brac{\int_{\mathbb{R}}\frac{dt}{2\pi}\abs{u_2+it}^{-2}}\mathbbm{1}+\dots}
(the corrections in $(z_2-\bar{z}_2)$ as well as the contributions of primary fields $\Psi_{b^{-2}/2}$ and $\Psi_{-b^{-2}/2-1}$ are now contained in the dots). The occuring integral is easily calculated to be
\eq{\int_{\mathbb{R}}\frac{dt}{2\pi}\abs{u_2+it}^{-2}=\abs{u_2+\bar{u}_2}^{-1}\,,}
so that again the asymptotics of $\tilde{{\cal F}}^{s}_{j,+}$ in (\ref{PlusLim}) is matched precisely. The $b^{-2}/2$-shift equation we obtain for the redefined one point amplitude (\ref{ReDef}) then reads
\eq{\label{ContShiftEq}\nu_{b}^{b^{-2}/2}(1+b^2)\tilde{c}(b^{-2}/2,0\arrowvert\alpha)f_{\sigma}(j)=f_{\sigma}(j+b^{-2}/2)\,.}

\section{\label{Consistency}Consistency of Discrete and Continuous $\bf AdS_2$ D-Branes}
The discrete $AdS_2^{(d)}$ branes of \cite{Ribault:DiscreteAdS2} have one point amplitudes
\eq{f_{\sigma}(j\arrowvert m,n)=\frac{i\pi\sigma e^{i\pi m}}{\Gamma(-b^2)\sin[\pi nb^2]}e^{-i\pi\sigma(m-\frac{1}{2})(2j+1)}\frac{\sin[\pi nb^2(2j+1)]}{\sin[\pi b^2(2j+1)]}\,,}
with $n,m\in\mathbb{Z}$. It is absolutely straightforward to check that they satisfy the $b^{-2}/2$-shift equation (\ref{DiscreteShiftEq}). Note that checking the $1/2$-shift equation, we actually only need $m\in\mathbb{Z}$. The additional restriction $n\in\mathbb{Z}$ is required when checking our novel $b^{-2}/2$-shift equation (\ref{DiscreteShiftEq}). The above amplitudes also satisfy the reflection symmetry constraint (\ref{RefSymmCnstr}), a fact that has of course already been checked in \cite{Ribault:DiscreteAdS2}.

Let us now turn our attention to the continuous $AdS_2^{(c)}$ branes of \cite{PST}. Their one point amplitudes read
\eq{f_{\sigma}(j\arrowvert\alpha)=-\frac{\pi A_b}{\sqrt{\nu_b}}\frac{e^{-\alpha(2j+1)\sigma}}{\sin[\pi b^2(2j+1)]}\,,}
with $\alpha\in\mathbb{R}$. Plugging that into the appropriate $b^{-2}/2$-shift equation (\ref{ContShiftEq}), we can infer an expression for the residuum of the bulk-boundary OPE coefficient
\eq{\tilde{c}(b^{-2}/2,0\arrowvert\alpha)=-\frac{e^{-\alpha\sigma b^{-2}}}{\nu_{b}^{b^{-2}/2}(1+b^2)}\,.}
This result should be compared to \cite{HosomichiRibault:SolutionOnDisc}, where general expressions for bulk-boundary coefficients have been given.

\section{Conclusion} We have shown that making use of the continuity axiom proposed in
\cite{HosomichiRibault:SolutionOnDisc}, the following desired facts about the $\rm H_3^+$
boundary CFT can be established: 
\begin{itemize} 
\item The two point function can be defined everywhere in the $(u,z)$ unit square (see equations (\ref{ConfBlocks1+}), (\ref{ConfBlocks1-}), (\ref{ConfBlocks1x}), (\ref{LinComb}) and (\ref{ConfBlocks2-}), (\ref{ConfBlocks2x}), (\ref{LinComb2+}), (\ref{LinComb2-x})). 
\item Factorization limits can be taken, resulting in the novel $b^{-2}/2$-shift equations (\ref{DiscreteShiftEq}) and (\ref{ContShiftEq}). They supplement the formlery known $1/2$-shift equations. 
\item The known discrete \cite{Ribault:DiscreteAdS2} and continuous \cite{PST} $AdS_2$ branes are shown to be consistent with these new constraints. For the discrete branes, that are labelled by a pair of parameters $(m,n)$, our new constraint additionally enforces $n\in\mathbb{Z}$ (the $1/2$-shift equation only restricts the parameters to $m\in\mathbb{Z}$). This fits in very nicely with Cardy's analysis \cite{Cardy:BCsFusionAndVerlinde} and associates the discrete $AdS_2$ branes 
to the  degenerate $\hat{\rm sl}(2,\mathbb{C})_k$ representations with $j_{m,n}:=-\frac{1}{2}+\frac{m}{2}+\frac{n}{2}b^{-2}$.
\end{itemize} 
Yet there is still one price to pay: The two point function in the patch $u<z$ is
not uniquely defined in the conformal blocks $\tilde{{\cal F}}^{s}_{j,-}$ and $\tilde{{\cal
F}}^{s}_{j,\times}$. This is the weakening of the Cardy-Lewellen constraints anticipated in
\cite{HosomichiRibault:SolutionOnDisc}. For our purpose of deriving $b^{-2}/2$-shift equations it
is however of no importance, because only the conformal block $\tilde{{\cal F}}^{s}_{j,+}$
contributes. These results are in total agreement with  \cite{HosomichiRibault:SolutionOnDisc}
and very strongly support their suggestion, that a proper definition of the $\rm H_3^+$ boundary
CFT has to include a continuity axiom.

\acknowledgments We like to thank Sylvain Ribault for drawing our attention to this problem.
Parts of H.A.'s work have been financially supported by the DFG-Graduiertenkolleg No. 282.

\appendix \section{Some Useful Formulae} \subsection{\label{Pochhammer}Pochhammer Symbol
Identities} The Pochhammer symbol is defined to be
\eq{(\alpha)_m:=\frac{\Gamma(\alpha+m)}{\Gamma(\alpha)}\,.} From this definition and the
functional equation of Euler's gamma function, $\alpha\Gamma(\alpha)=\Gamma(\alpha+1)$, one
easily derives the following identites: \spliteq{(\alpha)_{-m}&=&\frac{(-)^m}{(1-\alpha)_m}\,,\\
(\alpha)_{m+n}&=&\left\{\begin{array}{l} (\alpha+m)_n (\alpha)_m\\ (\alpha+n)_m (\alpha)_n
\end{array}\right.,\\ (\alpha)_{m-n}&=&\left\{\begin{array}{l} (\alpha+m)_{-n} (\alpha)_m\\
(\alpha-n)_m (\alpha)_{-n} \end{array}\right..}

\subsection{\label{AppellAndHorn}Appell's Function $\bf F_1$ and Horn's Function $\bf G_2$}
\paragraph{Definition as Convergent Series:} \noindent The definition of Appell's function $F_1$
is \eq{F_1(\alpha,\beta,\beta';\gamma\arrowvert
u;z):=\sum_{m,n=0}^{\infty}\frac{(\alpha)_{m+n}(\beta)_m(\beta')_n}{(\gamma)_{m+n}}\frac{u^m}{m!}
\frac{z^n}{n!}\,.} It is convergent for complex $u$ and $z$ in the domain $\abs{u}<1$,
$\abs{z}<1$. Clearly, for the third parameter $\gamma$ we need $\gamma\neq 0,-1,-2,\dots$. Horn's
function $G_2$ is defined by \eq{G_2(\beta,\beta';\alpha,\alpha'\arrowvert
u;z):=\sum_{m,n=0}^{\infty}(\beta)_m(\beta')_n(\alpha)_{n-m}(\alpha')_{m-n}\frac{u^m}{m!}\frac{z^
n}{n!}\,.} This series also converges for complex $u$ and $z$ with $\abs{u}<1$, $\abs{z}<1$. Its
parameters $\alpha$ and $\alpha'$ must be such that $\alpha\neq 1,2,3,\dots$ and $\alpha'\neq
1,2,3,\dots$. Both special functions are solutions to a certain system of partial differential
equations (see e.g. \cite{Exton}). This can be used to extend their definitions to domains
reaching outside $\abs{u}<1$, $\abs{z}<1$.

\paragraph{Generalized Series Representations:} \noindent Employing the Pochhammer symbol
identites stated in \ref{Pochhammer}, one deduces easily that
\eq{F_1(\alpha,\beta,\beta';\gamma\arrowvert u;z)=\sum_{n=0}^{\infty}\frac{(\alpha)_n
(\beta')_n}{(\gamma)_n}F(\alpha+n,\beta;\gamma+n\arrowvert u)\frac{z^n}{n!}\,,} $F$ being the
standard hypergeometric function. Of course, there is an analogous statment about the expansion
in the variable $u$. It is simply obtained by exchanging $\beta$ and $\beta'$ on the RHS.

The corresponding expansion for $G_2$ is obtained in the same manner and reads
\eq{G_2(\beta,\beta';\alpha,\alpha'\arrowvert u;z)=\sum_{n=0}^{\infty}\frac{(\alpha)_n
(\beta')_n}{(1-\alpha')_n}F(\alpha'-n,\beta;1-\alpha-n\arrowvert -u)\frac{(-z)^n}{n!}\,.} The
analogous expansion in the variable $u$ is of course obtained by exchanging $\alpha$ and
$\alpha'$ as well as $\beta$ and $\beta'$ on the RHS.

One should notice that for $\alpha\in\mathbb{Z}_{\leq 0}$, the above expansion breaks down,
because some of the occuring hypergeometric functions cease to be well defined (for
$\alpha\in\mathbb{Z}_{>0}$ the function $G_2$ is not defined anyway). For our purposes, the case
$\alpha=0$ becomes important when taking $u=z$ in (\ref{ConfBlocks2x}). In this case, it is
however not difficult to derive a similar expansion: \eq{G_2(\beta,\beta';0,\alpha'\arrowvert
u;z)=\sum_{n=0}^{\infty}\frac{(\beta)_n (\beta')_n}{(1)_n}F(\beta+n,\alpha';1+n\arrowvert
-u)\frac{(u\cdot z)^n}{n!}\,.}

\subsection{\label{OPE}OPE Coefficients} OPE coefficients are derived from the structure
constants that were given in \cite{Teschner:OPE}. It is important to take into consideration the
different normalizations of field operators. In \cite{Teschner:OPE}, the operators $\phi_j(u|z)$
are used, whereas here (as well as in \cite{PST}) we are dealing with
$\Theta_j(u|z):=B^{-1}(j)\phi_j(u|z)$, with $B(j)=(2j+1)R(j)/\pi$, and $R(j)$ the reflection
amplitude (\ref{Rj}). Accordingly, the structure constants $D(j,j_1,j_2)$ of \cite{Teschner:OPE}
have to be multiplied by some factors of $B^{-1}$ in order to extracxt the expressions relevant
for our conventions: \eq{C(j,j_1,j_2):=D(j,j_1,j_2)B^{-1}(j_1)B^{-1}(j_2)\,.} Now, the singular
vector labelled by $b^{-2}/2$ restricts the possibly occuring field operators in the operator
product to those with labels $j_+:=j+b^{-2}/2$, $j_-:=j-b^{-2}/2$ and
$j_{\times}:=-j-1-b^{-2}/2$. The OPE therefore reads \aligneq{\Theta_{b^{-2}/2}(u_2\arrowvert
z_2)\Theta_j(u_1\arrowvert
z_1)&\simeq\sum_{\epsilon=+,-,\times}\abs{z_2-z_1}^{-2[h(b^{-2}/2)+h(j)-h(j_{\epsilon})]}\times\qquad\nonumber\\
&\qquad\times\abs{u_2-u_1}^{2[b^{-2}/2+j-j_{\epsilon}]}C_{\epsilon}(j)\Theta_{j_{\epsilon}}(u_1\arrowvert z_1)\,.} For the corresponding OPE coefficients, we calculate \begin{gather}
C_+(j)=1\,,\\ C_-(j)=-\nu_b^{-b^{-2}}\ebrac{b^2(2j+1)}^{-2}\,,\\
C_{\times}(j)=-\frac{\nu_b^{-2j-1-b^{-2}}}{b^4}\frac{\Gamma(1+b^{-2})}{\Gamma(1-b^{-2})}\frac{\Gamma(1+2j)\Gamma(-1-2j-b^{-2})\Gamma(-b^2(2j+1))}{\Gamma(-2j)\Gamma(2+2j+b^{-2})\Gamma(1+b^2(2j+
1))}\,. \end{gather}

\bibliographystyle{utphys} \bibliography{bibfile}

\end{document}